\documentclass[a4paper, amsfonts, amssymb, amsmath, reprint, showkeys, nofootinbib, twoside]{revtex4-1}
\usepackage[english]{babel}
\usepackage[utf8]{inputenc}
\usepackage[colorinlistoftodos, color=green!40, prependcaption]{todonotes}
\usepackage{amsthm}
\usepackage{mathtools}
\usepackage{physics}
\usepackage{xcolor}
\usepackage{graphicx}
\usepackage[left=23mm,right=13mm,top=35mm,columnsep=15pt]{geometry} 
\usepackage{adjustbox}
\usepackage{placeins}
\usepackage[T1]{fontenc}
\usepackage{lipsum}
\usepackage{csquotes}
\usepackage[pdftex, pdftitle={Article}, pdfauthor={Author}]{hyperref} 
\usepackage{amssymb}
\usepackage{natbib}

\usepackage[LGRgreek]{mathastext} 
\newcommand\tab[1][1cm]{\hspace*{#1}} 

\usepackage{lineno}
\usepackage{setspace}
\usepackage{hyperref}

\begin{document}
\title{Sol-gel synthesized double perovskite Gd$_2$FeCrO$_6$ nanoparticles : Structural, magnetic and optical properties}

\author{M. D. I. Bhuyan, Subrata Das and M. A. Basith}
    \email[Email address: ]{mabasith@phy.buet.ac.bd}
    \affiliation{Nanotechnology Research Laboratory, Department of Physics, Bangladesh University of Engineering and Technology, Dhaka-1000, Bangladesh.\\ \\ DOI: \href{https://doi.org/10.1016/j.jallcom.2021.160389}{10.1016/j.jallcom.2021.160389}}


\begin{abstract}
Lead-free double perovskites are overtaking single perovskites as solar harvesting materials due to their superior stability, excellent catalytic efficiency and minimal toxicity. In this investigation, we have synthesized double perovskite Gd$_2$FeCrO$_6$ (GFCO) nanoparticles for the first time via a facile sol-gel technique to investigate their structural, magnetic and optical properties. The double perovskite GFCO crystallized in monoclinic structure with P2$_1$/n space group. The Fe/Cr-O bond length was calculated as $\sim$1.95 \AA $\;$ from the Raman spectrum which was consistent with the value, $\sim$1.99 \AA $\;$ obtained from X-ray diffraction analysis. The average size of the nanoparticles was determined to be $\sim$70 nm by both field emission scanning electron microscopy and transmission electron microscopy. The existence of mixed valence states of Fe and Cr was confirmed by X-ray photoelectron spectroscopy. The zero field cooled (ZFC) and field cooled (FC) curves largely diverged below 20 K. A downturn was observed in the ZFC curve at 15 K which
corresponds to an antiferromagnetic, N\'eel transition. The narrow magnetic hysteresis loop recorded at 5 K was nearly saturated and demonstrated an asymmetric shift along the magnetic field axis indicating the concurrence of ferromagnetic and antiferromagnetic domains in GFCO nanoparticles. The UV-visible and photoluminescence spectroscopic analyses unveiled the semiconducting nature of nanostructured GFCO with an optical band gap of 2.0 eV. The as-synthesized thermally stable lead-free GFCO semiconductor might be a potential perovskite material to be employed in photocatalytic and related solar energy applications due to its ability to absorb the visible spectrum of the solar light efficiently. 
\end{abstract}


\maketitle

\section{Introduction}

In the last few decades, perovskite oxides, ABO$_3$ (A = rare earth metal ions, B = transition metal ions) have intrigued significant research interest due to their exceptional structural and compositional flexibility \cite{lv2019copper/cobalt,cohen1992origin,bhalla2000perovskite,basith2017preparation,basith2014room}. Recently, substituting exactly half of the transition metal ions of perovskites with another cations has resulted in a number of interesting double perovskites, A$_2$BB$^{\prime}$O$_6$ ( e.g. A = La, Y, Pr, Nd, Gd, etc., B and B$^{\prime}$ = Fe, Cr, Ni, Mn, etc.) which possess a great diversity of functional properties based on their various cationic orderings and oxidation states \cite{vasala2015a2b,gray2010local,das2017pr2fecro6,das2008electronic,bull2003determination,yadav2015magnetic,costilla2020double}. Notably, double perovskites can be effectively employed as magnetic memory components owing to their interesting magnetic properties e.g. half metallicity, magnetic ordering, magnetoresistance and multiferroicity \cite{masud2015occurrence,du2010magnetic}. Therefore, previously the magnetic properties of a number of double perovskites were investigated both experimentally as well as theoretically for their potential applications in the next generation spintronic devices \cite{palakkal2015observation,basith2007temperature,das2019enhanced,kobayashi1998room}. Recently, it has been demonstrated that the optical and electronic properties of this class of double perovskite materials can be tuned by semiconductor engineering to use them in many advanced technological fields such as photocatalysis, photovoltaic devices, photo(electro)chemical energy storage systems etc \cite{lv2016structural,idris2020novel,halder2019investigating,shirazi20202}.

It is noteworthy that among different double perovskites, A$_2$FeCrO$_6$ (where A = La, Pr, Bi) based compounds have demonstrated fascinating magnetic properties because of possessing two 3d transition elements i.e. Fe and Cr in their crystal structure  \cite{ das2017pr2fecro6, miura2001electronic, wu2020structural}. Based on Goodenough-Kanamori rules, these materials are predicted to have ferromagnetic order owing to the 180$^{\circ}$ superexchange interaction between Fe$^{3+}$ and Cr$^{3+}$ ions via oxygen \cite{kanamori1959superexchange,goodenough1955theory}. For instance, ferromagnetic spin order has been observed in ordered double perovskite \cite{miura2001electronic, ueda1998ferromagnetism}, La$_2$FeCrO$_6$ whereas their single-phase materials LaCrO$_3$ \cite{weinberg1961electron} and LaFeO$_3$ \cite{peterlin1986antiferromagnetic} possess antiferromagnetic (AFM) state. A recent investigation has reported that Pr$_2$FeCrO$_6$ double perovskite exhibits room-temperature ferromagnetic ordering with high Curie temperature as well as weak ferroelectricity \cite{ravi2018multiferroism}. Interestingly, unlike La$_2$FeCrO$_6$ and Pr$_2$FeCrO$_6$ double perovskites, Bi$_2$FeCrO$_6$ has exhibited weak ferromagnetic (FM) behavior under a strong antiferromagnetic background \cite{wu2020structural}. Moreover, in recent times, Bi$_2$FeCrO$_6$ and Pr$_2$FeCrO$_6$ double perovskites have manifested fascinating optoelectronic properties such as favorable band gap energy, strong absorbance in the visible range of the solar spectrum and notable efficiency in photocatalysis \cite{vasala2015a2b,wu2020structural,gaikwad2019structural}. Such diversified magnetic and optical properties intrigue significant research interest to explore other members of A$_2$FeCrO$_6$ double perovskite family.

Nevertheless, it is difficult to form ordered structure of A$_2$FeCrO$_6$ double perovskites because of the almost equal ionic radii of Fe and Cr ions \cite{gray2010local}. The complementary ionic radii of these two ions result in antisite disorder and break the ideal alternating order of Fe and Cr \cite{gray2010local,miura2001electronic}. The physical properties of these material systems are strongly related to the variation of antisite disorder \cite{gray2010local}. For instance, a ferromagnetic state was observed in La$_2$FeCrO$_6$ perovskite, however, the measured saturation magnetic moment was significantly lower than the theoretical value \cite{ueda1998ferromagnetism}. Such discrepancy can be attributed to the fact that the condition for the synthesis of the sample was not up to the mark and the degree of order between Fe and Cr was not precisely considered to analyze magnetic properties. Due to the difficulties inherent in the formation of crystal structure controlling the order-disorder effects, most of the double perovskites of A$_2$FeCrO$_6$ family are yet less explored. To the best of our knowledge, synthesis and characterization of nanostructured Gd$_2$FeCrO$_6$ (GFCO) double perovskite have not been reported yet. Therefore, in the current investigation, we have synthesized GFCO nanoparticles by adapting a low-cost facile sol-gel method \cite{iranmanesh2016sol} and investigated their crystallographic structure, chemical composition along with magnetic and optical properties extensively. Finally, the potential applicability of this newly synthesized material in the field of photocatalysis and oxygen evolution was discussed in view of its favorable band gap as well as band edge position.
    
\section{EXPERIMENTAL DETAILS}

Double perovskite GFCO was prepared by adopting a citrate-based sol-gel technique \cite{cernea2013preparation,bijelic2020rational,hasan2016saturation}. Stoichiometric amounts of analytical grade Gd(NO$_3$)$_3$.6H$_2$O (Sigma-Aldrich, 99.9\%), Fe(NO$_3$)$_3$.9H$_2$O (Sigma-Aldrich, 99.95\%) and Cr(NO$_3$)$_3$.9H$_2$O (Sigma-Aldrich, 99\%) were dissolved separately in 40 ml of deionized water using a magnetic stirrer. Afterward, the solutions were mixed and then, citric acid (CA) and ethylene glycol (EG) were added to this mixture ensuring the molar ratio of (Gd$^{3+}$, Fe$^{3+}$+ Cr$^{3+}$) : (CA) : (EG) = 1 : 1 : 4 to develop a polymeric-metal cation network \cite{chanda2015structural}. Then, the solution was stirred and heated at 80 $^{\circ}$C for around 4 h. Gel precursor was obtained and then burnt at elevated temperature to yield brown fluffy powder. Thereafter, the powder was ground by a mortar and calcined at 800 $^{\circ}$C for 6 h in air followed by cooling down to room temperature (RT).

The crystallographic structure of as-synthesized GFCO double perovskite was investigated by collecting their powder X-ray diffraction pattern (XRD) using a diffractometer (PANalytical Empyrean) with a Cu X-ray source (wavelength, $\lambda$: K$_{\alpha 1}$ = 1.540598 \AA $\;$ and K$_{\alpha2}$ = 1.544426 \AA).  Rietveld refinement of the obtained XRD data was performed by the FullProf computer program package \cite{rodriguez1990fullprof} in order to determine the structural parameters. The Raman spectra was obtained in backward scattering configuration with 532 nm laser excitation by a MonoVista CRS+ Raman microscope systems (S\&I). Fourier Transform Infrared (FTIR) spectrum of GFCO was collected using a Spectrum Two FT-IR Spectrometer (PerkinElmer). The transmittance value of the sample was measured for infrared (IR) radiation with wavenumbers ranging from 350 to 4000 cm$^{-1}$. Thermal analysis of the prepared powder materials of GFCO was performed by Thermogravimetric analysis (TGA) measurement (NETZSCH, STA 449 F3 Jupiter) in nitrogen atmosphere at a heating rate of 10 $^{\circ}$C/min from 30 $^{\circ}$C to 1000 $^{\circ}$C. Field emission scanning electron microscope (FESEM) (XL30SFEG; Philips, Netherlands and S4300; HITACHI, Japan) was used to analyze the surface morphology of GFCO perovskite. The elemental composition of as-synthesized GFCO was examined by performing energy-dispersive X-ray (EDX) spectroscopy using an X-ray spectroscope. For a further investigation of the size and crystallinity of the synthesized GFCO particles, the transmission electron microscopy (TEM) (Talos F200X) imaging was carried out. X-ray photoelectron spectroscopy (XPS) has been performed using a PHI5600 ESCA 120 (ULVAC PHI) system to determine the chemical binding energies and valence states of different elements in the compound. To extract XPS spectra monochromatized Al K$_\alpha$ X-ray (1486.6 eV) and the 121 hemispherical mirror analyzer with a total energy resolution of $\sim$0.13 eV and the 122 emission angle of 45° were used. The background of XPS spectra was subtracted by the Shirley 123 procedure and the peaks were fitted using the Gaussian-Lorentzian function. Temperature dependent magnetization measurement of the as-prepared sample was performed using a Quantum Design Physical Property Measurement System (PPMS) under 100 Oe applied magnetic field via zero-field cooled (ZFC) and field cooled (FC) methods \cite{basith2015tunable}. Further, field dependent magnetization measurements of GFCO were conducted at 300, 200, 100 and 5 K using the same PPMS. An ultraviolet-visible (UV-visible) spectrophotometer (UV-2600, Shimadzu) was used to obtain absorbance spectrum of the as-synthesized GFCO for wavelengths ranging from 200 to 800 nm. Steady-state photoluminescence (PL) spectroscopy was carried out at room temperature using a Spectro Fluorophotometer (RF-6000, Shimadzu).

\section{RESULTS AND DISCUSSIONS}
\subsection{Crystallographic analysis}
\begin{figure*}[t]
\centering
\includegraphics[width=160 mm, scale=0.8]{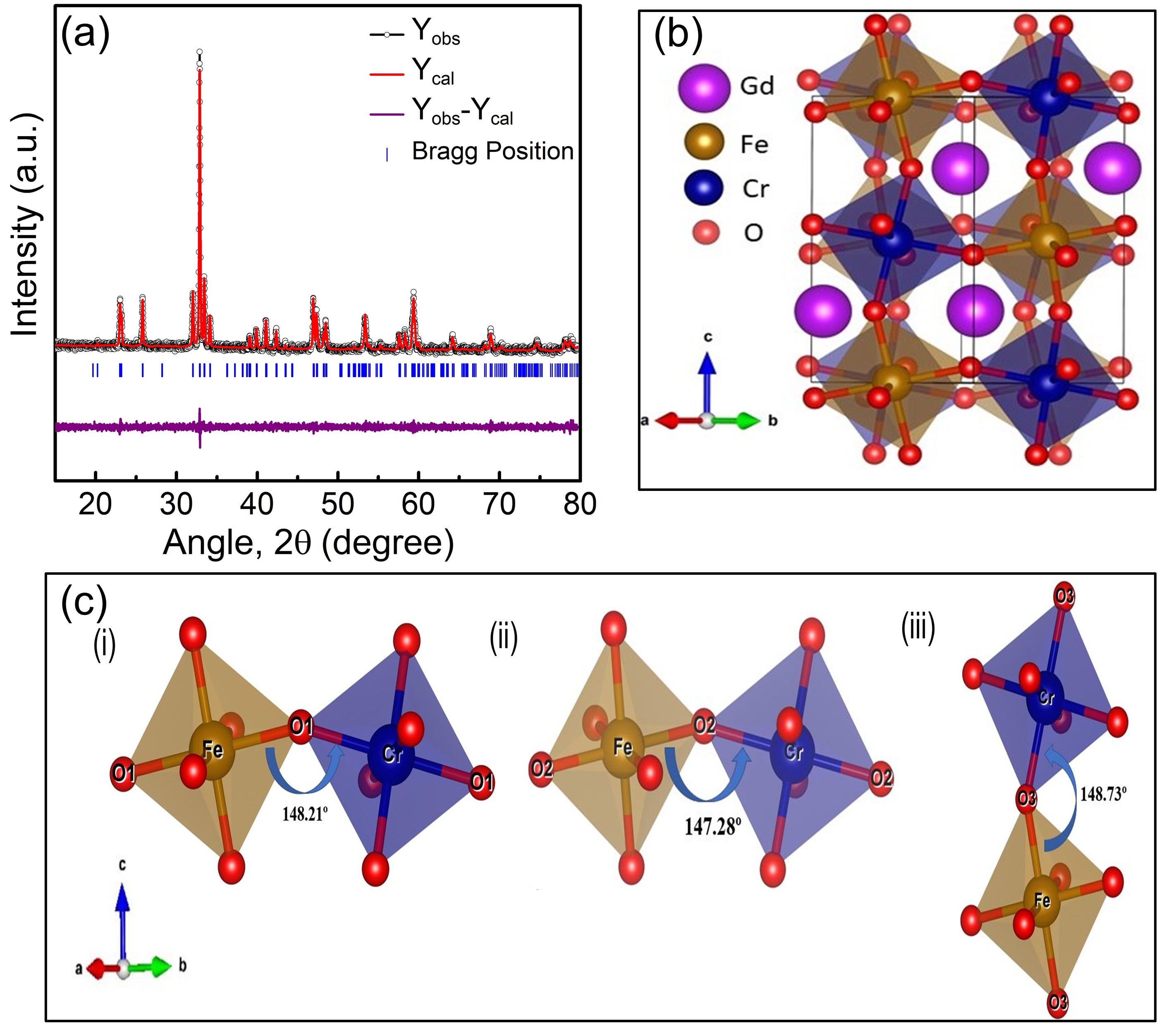}
\caption{\label{fig:epsart} (a) Rietveld refined powder XRD spectrum of Gd$_2$FeCrO$_6$ nanoparticles at room temperature. (b) Schematic representation of the Gd$_2$FeCrO$_6$ monoclinic unit cell. (c) Magnified view of interconnected Fe/CrO$_6$ octahedra (i), (ii) in the ab-plane and (iii) along c-axis.}
\end{figure*}

\begin{table*}[t]

\centering
\caption{\label{tab:table3}Structural parameters of Gd$_2$FeCrO$_6$ double perovskite as obtained via Rietveld refinement of the XRD pattern and reliability (R) factors.}

\begin{tabular}{lllllllllll}
\hline
Atom & Wyc. & $x$ & $y$& $z$& Bond length (\AA) & Bond angle ($^{\circ}$) & Tilt angle ($^{\circ}$) & R factors\\
& positions &  & & & & & & & \\\hline 
 Gd&$4e$&${0.513}$&${0.560}$&${0.252}$&$Fe-O_1=2.00$
 &$<Fe-O_1-Cr>$&$ {15.90}$&R$_{exp}=4.18$ \\
  & & & & & &$={148.2}$& & \\
 Fe&$2c$&$0$&$0.5$&$0$&$Fe-O_2=1.97$
 &$<Fe-O_2-Cr>$& ${16.35}$&$R_{wp}=3.93$\\
   & & & & & &$={147.3}$& & \\
 Cr&$2d$&$0.5$&$0$&$0$&$Fe-O_3 = 1.99$
 &$<Fe-O_3-Cr>$& ${15.65}$&$\chi^2=1.13$\\
   & & & & & & $= {148.7}$& & \\
 O$_1$&$4e$&${0.207}$&${0.207}$&${-0.055}$&$Cr-O_1 = 2.00$\\
 O$_2$&$4e$&${0.299}$&${0.693}$&${-0.051}$&$Cr-O_2 = 2.02$\\
 O$_3$&$4e$&${0.402}$&${-0.019}$&${0.249}$&$Cr- O_3 = 1.98$\\
 \hline
\end{tabular}

\end{table*}
Initially, we had calculated the tolerance factor and global instability index of GFCO using the SPuDS- Structure Prediction Diagnostic Software to get an idea about its crystal structure \cite{lufaso2001prediction}. The tolerance factor ($t$) of double perovskite GFCO can be expressed as \cite{anderson1993b,li2018new},

\begin{eqnarray}
 t=\frac{r_{Gd}+r_O}{\sqrt{2(\frac{r_{Fe}+r_{Cr}}{2}+r_O)}}
\end{eqnarray} 
Here, $r_{Gd}$, $r_{Fe}$, $r_{Cr}$ and $r_O$ denote the ionic radii of Gd, Fe, Cr cations and O anion, respectively. Generally, for double perovskites, the value of $t\approx1$ corresponds to ideal cubic structure. If $t$ is greater than 1, the crystal structure would be hexagonal and if $t < 0.97$, the structure can be either orthorhombic or monoclinic \cite{anderson1993b}. Notably, the $t$ of GFCO double perovskite was calculated to be 0.91 at room temperature (RT) from which its crystal structure can be predicted as either orthorhombic or monoclinic.

It should be noted that the degree of stability of a certain phase of perovskite can be estimated by calculating its global instability index (GII) which is the difference between the calculated bond-valence sum (BVS) and the formal valence of cations and anions. GII is defined as the root mean square of bond discrepancy factor in the unit cell \cite{yamada2018complementary, ravi2017room}:
\begin{eqnarray}
 GII=\sqrt{\frac{\sum_{i=1}^{N}(d_i)^2}{N}}
\end{eqnarray} 
where $d_{i}$ is the bond discrepancy factor calculated from bond valence sum and $N$ is the number of ions. Typically, for unstrained structures, the GII values are smaller than 0.1 valence unit (v. u.) \cite{ravi2017room}. Especially, as reported by a survey of literature \cite{li2018new, byeon2003high}, stable double perovskite crystal structures can be formed under ambient pressure which have GII value smaller than 0.02 v. u \cite{li2018new, byeon2003high}. Notably, the GII value of GFCO is found to be 0.013 v. u. which is conspicuously smaller than 0.02 v. u. As a consequence, even without employing high-pressure during the synthesis, we achieved unstrained single phase crystal structure of GFCO.

Fig. 1(a) demonstrates the Rietveld refined powder XRD spectrum of as-prepared GFCO double perovskite at RT \cite{mcsucker1999p, rietveld1969profile}. Notably, no undesired secondary phase was detected in the XRD spectrum which confirms the high phase purity of the sample. Since the value of $t$ is $<0.97$, hence for Rietveld refinement, we have considered monoclinic P2$_1$/n space group as the starting model. The lattice parameters, $a$, $b$ and $c$ of this monoclinic symmetry, are related to the cell parameter, $a_0$ $\approx$ 3.8 \AA, of ideal cubic perovskite, as $a$ $\approx$ $b$ $\approx$ $\sqrt{2}a_0$, and $c$  $\approx$ 2$a_0$. The background was modeled by linear interpolation between a set of background points with refinable heights. Bragg’s reflections were modeled using Thompson-Cox-Hastings pseudo-Voigt Axial divergence asymmetry. Initially, cell parameters and scale factor were refined followed by the refinement of profile and full width at half maximum (FWHM) parameters. After achieving proper profile matching, the thermal and positional coordinates were refined. We have also considered orthorhombic crystal structure with Pbnm space group as the model for refinement. Between these two models, better fitting was obtained for monoclinic structure with P2$_1$/n space group. However, since Fe and Cr possess almost the same scattering factor, it is difficult to confirm only by XRD technique whether GFCO is ordered or not.
Yet one can obtain information about cation ordering by checking the superstructure reflection around 20$^{\circ}$
\cite{shi2011local, chanda2016magnetic}. As shown in Fig. 1, the superstructure peak of GFCO at 20.2$^{\circ}$ is very weak which suggests that the B-site cation ordering was not perfect in the synthesized sample. An indication of the absence of proper long-range ordering was also evident from the mixed valence states of Fe and Cr obtained by XPS analysis as will be discussed later on.

From the Rietveld refinement, the cell parameters of GFCO double perovskite are found to be $a$ = 5.359(1) \AA, $b$ = 5.590(2) \AA, $c$ = 7.675(3) \AA, $\alpha = \gamma$ = 90$^{\circ}$ and $\beta$ = 89.96(1)$^{\circ}$ with cell volume 229.9 $\AA^3$ which are in well agreement with those of analogous double perovskite materials \cite{das2017pr2fecro6,yadav2015magnetic,masud2015occurrence}. Using these parameters, we have modeled the unit cell of GFCO by the Visualization for Electronic and Structural Analysis (VESTA) software \cite{momma2011vesta}. The monoclinic unit cell of GFCO with corner shared Fe/CrO$_6$ octahedra is shown in Fig. 1(b). It can be observed that the Fe and Cr cations are not exactly at the center of octahedra indicating octahedral distortion. For a further insight, we have presented the enlarged view of the three interconnected Fe/CrO$_6$ octahedra in the ab plane and along c-axis in Fig. 1(c). To estimate the distortion, we have also estimated the tilt angle ($\Phi$) which is defined as $\Phi$ = (180-$\theta$)/2, where $\theta$ is the Fe/Cr-O-Fe/Cr bond angle \cite{das2017pr2fecro6}. The average value of $\Phi$ is found to be 15.96$^{\circ}$ which further confirms the structural distortion of GFCO unit cell. Table I summarizes other structural parameters i.e. Wyc. positions, atomic coordinates, bond lengths and bond angles of synthesized GFCO double perovskite obtained by Rietveld refinement as well as the reliability (R) factors. Notably, the bond distances and bond angles shown in Table I agree quite well with the reported values in the octahedral oxygen environment of related double perovskites \cite{das2017pr2fecro6,masud2015occurrence}. 

\begin{figure*}[t]
\centering
\includegraphics[width=160mm, scale=0.8]{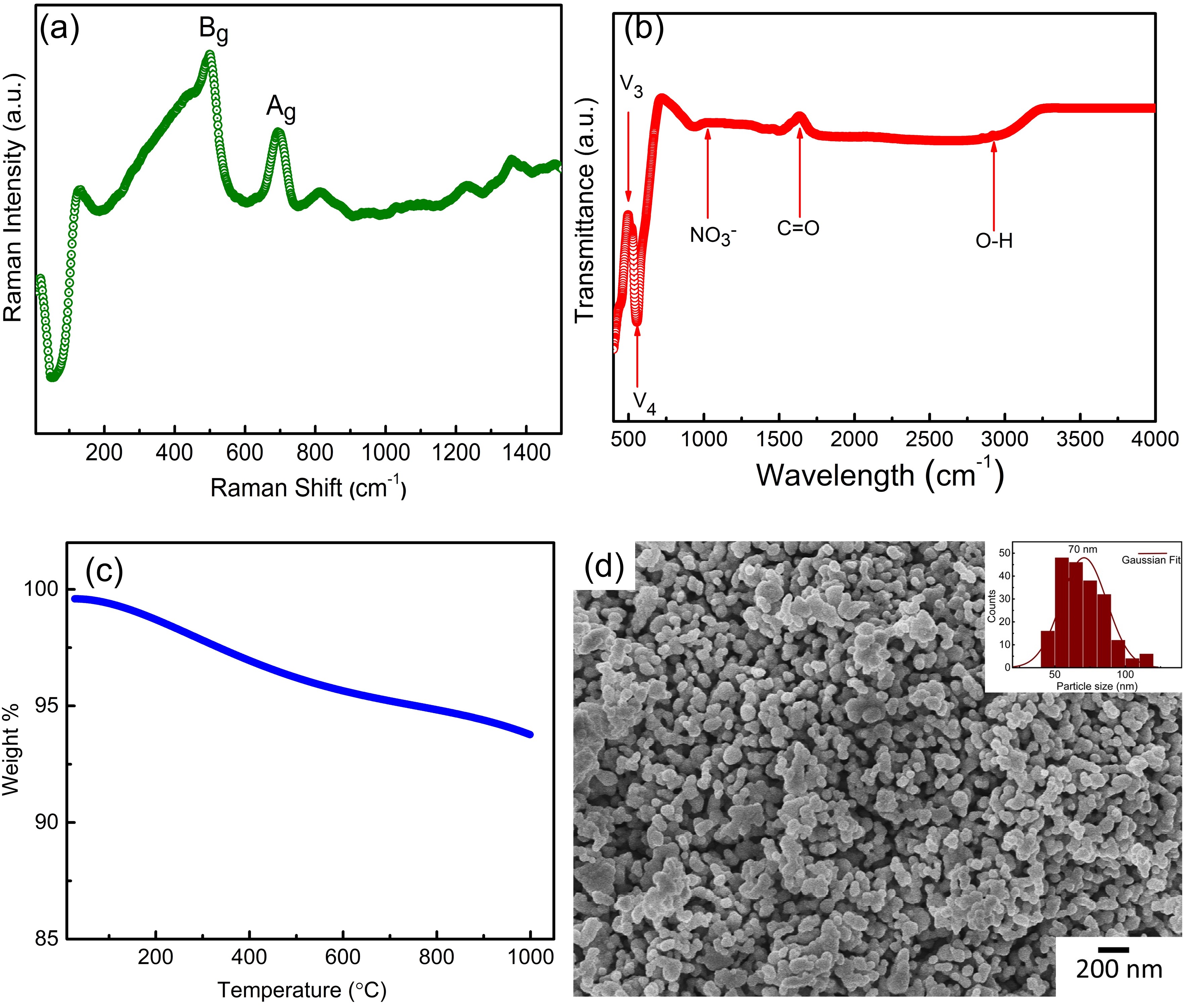}
\caption{\label{fig:epsart} (a) Raman and (b) FTIR spectra of Gd$_2$FeCrO$_6$ particles recorded at room temperature. (c) TGA curve of Gd$_2$FeCrO$_6$ powders from 30 $^{\circ}$C to 1000 $^{\circ}$C in N$_2$ with heating rate of 10 $^{\circ}$C/min. (d) FESEM image of Gd$_2$FeCrO$_6$ nanoparticles. Inset: particle size distribution histogram.}
\end{figure*}

Raman spectroscopy has been performed at RT to further investigate the crystal structure, cation ordering  and spin-phonon coupling of as-synthesized GFCO double perovskite [Fig. 2(a)]. Notably, the vibrational modes observed in the Raman spectrum arise from the Fe/CrO$_6$ octahedra and Gd-O bonds. Since the as-synthesized sample possesses monoclinic crystal structure with P2$_1$/n space group, two main vibrational modes, A$_g$ and B$_g$ are expected in the range of 500 to 700 cm$^{-1}$ in the Raman spectrum. The Raman spectrum of GFCO [Fig. 2(a)] displayed two broad modes near 511 cm$^{-1}$ and 690 cm$^{-1}$ which correspond to B$_g$ and A$_g$ modes, respectively and thus, confirm the monoclinic P2$_1$/n symmetry \cite{nasir2019role, guo2008influence, iliev2007raman}. Notably, A$_g$ mode arises due to stretching (breathing) vibrations of Fe/CrO$_6$ octahedra \cite{iliev2007raman} whereas B$_g$ mode occurs due to both anti-stretching and bending motions. Moreover, both the first-order strong modes i.e. A$_g$  and B$_g$ are fairly asymmetric. In these modes, the broadness and asymmetry might depend on several factors such as incomplete ordering of Fe and Cr sites, nearly same frequency of various Fe/Cr-O vibrations \cite{nasir2019role} etc.

Furthermore, multiband can be observed in Fig. 2(a) at the range of 800 to 1410 cm$^{-1}$ which is associated with multi-phonon scattering. It is noteworthy that modes around 820 cm$^{-1}$ and around 1355 cm$^{-1}$ are the overtones of 
B$_g$ and A$_g$, respectively. The mode at 1206 cm$^{-1}$ can be considered as a combination of both B$_g$ and A$_g$ modes \cite{guo2006growth}. A stretching mode can also be noticed in the low frequency regime (at around 150 cm$^{-1}$),  which provides further evidence for the monoclinic P2$_1$/n space group \cite{nasir2019role}. Such low frequency vibrations have arisen due to coupled Fe/CrO$_6$ tilting vibrations and Gd-O stretching. It is noteworthy that any intense mode around 377 cm$^{-1}$ is the indication of oxide impurity. For as-synthesized GFCO, the most intense Raman mode is observed around 511 cm$^{-1}$ which confirms the phase purity of the synthesized double perovskite and supports the result of XRD analysis \cite{panitz2000raman}. This intense mode can be further used to calculate the Fe/Cr-O bond length by using the following equation \cite{obregon2014improved}:
\begin{eqnarray}
 v(cm^{-1})=21349\;exp(-1.9176R \;(\AA))
\end{eqnarray}
where $v$ denotes the stretching Raman frequency and $R$ is the Fe/Cr-O bond length. The calculated bond length of Fe/Cr-O is  $\sim$1.95 $\AA$ which is consistent with the average bond length of Fe-O and Cr-O ($\sim$1.99 $\AA$) as obtained via Rietveld refinement.

FTIR analysis has been performed at RT to investigate the chemical structure of synthesized GFCO double perovskite as demonstrated in Fig. 2(b). Notably, the Fe/CrO$_6$ octahedra complex in the unit cell of GFCO is expected to have six normal modes of vibrations (V$_1$ – V$_6$). Among these six vibrations, V$_1$, V$_2$ and V$_5$ would be Raman active, V$_3$ and V$_4$ would be IR active and V$_6$ would be inactive in both \cite{gaikwad2019structural,nakamoto2006infrared}. In the FTIR spectrum of GFCO [Fig. 2(b)], the transmission mode at around 508 cm$^{-1}$ can be associated with V$_3$ which indicates the existence of bending vibrations of Fe/Cr–O bonds in the Fe/CrO$_6$ octahedra \cite{gaikwad2019structural,nakamoto2006infrared}. The broad transmission band V$_4$ centered at 555 cm$^{-1}$ corresponds to Fe/Cr–O stretching vibrations \cite{gaikwad2019structural,nakamoto2006infrared}. The weak transmission band near 930 cm$^{-1}$ can be attributed to the NO${_3}^{-}$ ions which are trapped in the particles \cite{nakamoto2006infrared}. The weak bands present at 1400 to 1600 cm$^{-1}$ exist owing to the vibrations of C–O bonds which arise because of the chemisorption of ambient CO$_2$ by the surface of GFCO \cite{gaikwad2019structural}. Finally, the IR bands in the range of 2840 to 3200 cm$^{-1}$ might be imputed to the stretching vibrations of O–H bonds due to the adsorption of water at the surface \cite{gaikwad2019structural}.

\subsection{Thermal stability and morphological analyses}

The thermal stability which is an important property for perovskite materials for their future applications was evaluated via TGA experiment of the as-prepared GFCO powder sample. The TGA curve shown in Fig. 2(c) demonstrated the great thermal stability of GFCO powders with a subsequent weight loss of only $\sim$6\% up to temperature 1000 $^{\circ}$C which is comparable to that of the some other double perovskites \cite{cowin2015conductivity,ding2016redox}. The initial weight loss at a temperature below 150 $^{\circ}$C may be due to the desorption of adsorbed water and gases. The weight loss at higher temperatures up to 1000 $^{\circ}$C was gradual which are probably due to the reduction of both Fe and Cr ions associated with the loss of lattice oxygen.  Notably, from 600 $^{\circ}$C to 1000 $^{\circ}$C, the weight loss is less than $\sim$2\%, indicating that the sample can be calcined within this temperature range. Therefore, considering the stability of the material over a wide range of temperature, it is worth noting that the annealing of the sample at 800 $^{\circ}$C was reasonable.

\begin{table}[t]
\centering
\caption{Mass and atomic percentages of Gd$_2$FeCrO$_6$ nanoparticles as obtained by EDX analysis.}
\resizebox{\columnwidth}{!}{%
\begin{tabular}{lllll}
\hline
Element & \begin{tabular}[c]{@{}l@{}}Mass (\%)\\ (Theoretical)\end{tabular} & \begin{tabular}[c]{@{}l@{}}Mass (\%)\\ (Experimental)\end{tabular} & \begin{tabular}[c]{@{}l@{}}Atom (\%)\\ (Theoretical)\end{tabular} & \begin{tabular}[c]{@{}l@{}}Atom (\%)\\ (Experimental)\end{tabular} \\ \hline 
Gd      & 60.67                                                             & 59.18                                                              & 20                                                                & 20.85                                                              \\
Fe      & 10.77                                                             & 13.74                                                               & 10                                                                & 13.63                                                             \\
Cr      & 10.03                                                             & 11.79                                                               & 10                                                                & 12.56                                                               \\
O       & 18.52                                                             & 15.29                                                              & 60                                                                & 52.96                                                             \\ 
Total   & 100                                                               & 100                                                                & 100                                                               & 100  \\
\hline
\end{tabular}
}
\end{table}

\begin{figure*}[]
\centering
\includegraphics[width= 120 mm]{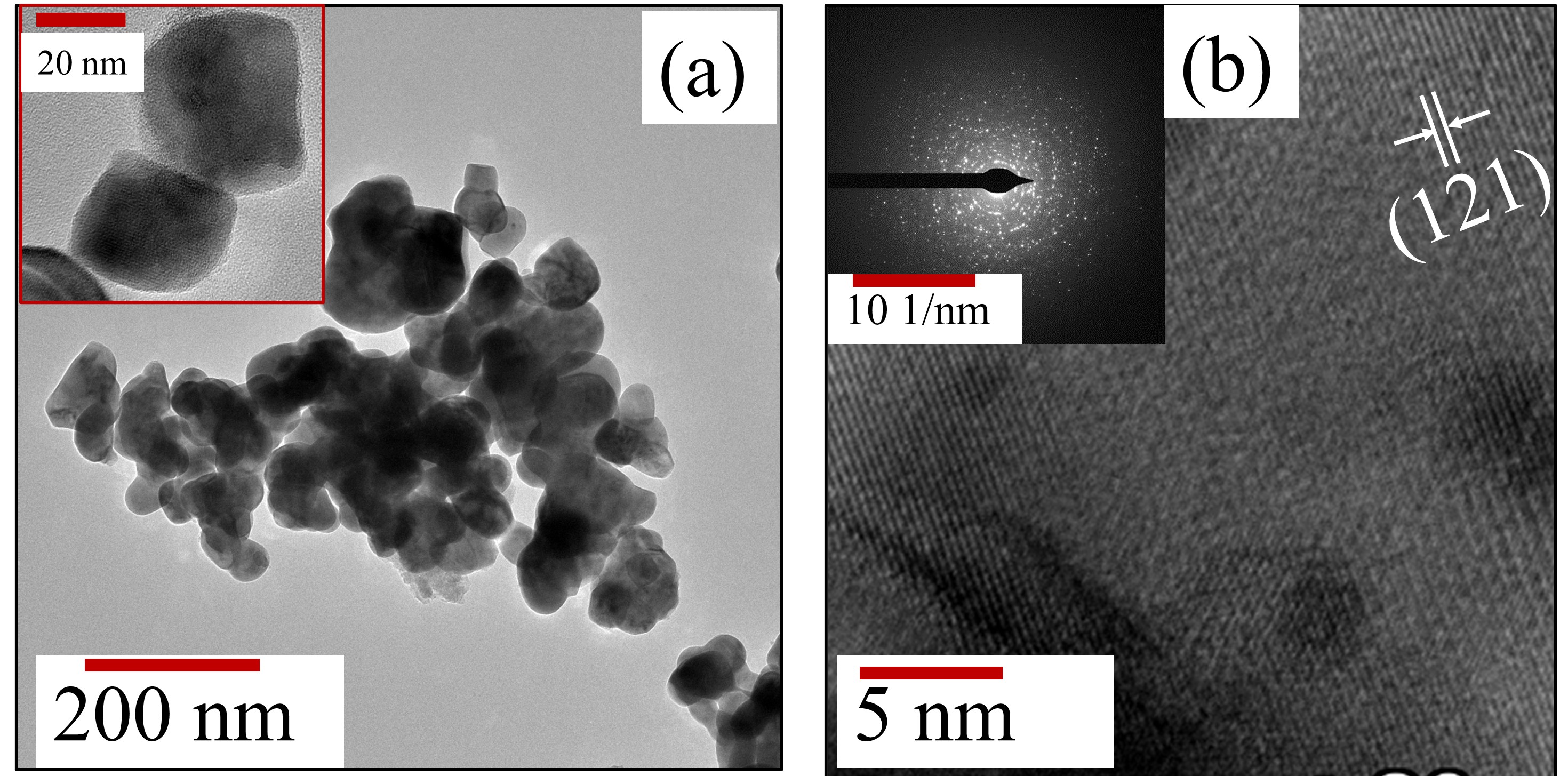}%
\caption{\label{fig:epsart}(a) Bright field TEM image of synthesized Gd$_2$FeCrO$_6$ double perovskite.  Inset of (a) shows individual nanoparticles. (b) HRTEM image of nanostructured Gd$_2$FeCrO$_6$ showing crystal planes. Inset shows SAED pattern of Gd$_2$FeCrO$_6$ nanocrystal.}
\end{figure*}
The FESEM image of GFCO double perovskite presented in Fig. 2(d) shows successful synthesis with a satisfactorily homogeneous surface. Notably, in a previous investigation \cite{gaikwad2019structural}, a similar class of double perovskite Pr$_2$FeCrO$_6$ was synthesized using wet-chemical technique co-precipitation. The authors observed significant agglomeration in the synthesized nanoparticles due to the higher rate of precipitation \cite{gaikwad2019structural}. However, in this present investigation, by using sol-gel technique we have synthesized GFCO particles which are not severely agglomerated. Inset of Fig. 2(d) displays the particle size distribution histogram as obtained from the FESEM image. From this figure, it can be estimated that the particle size of GFCO was mostly within the range of $\sim$40 to 100 nm with $\sim$70 nm average. Further, we have conducted EDX analysis at RT to evaluate the elemental composition of synthesized GFCO nanoparticles. The experimentally obtained percentages of mass and atom are provided in Table II. In order to compare, the theoretically calculated percentages of mass and atom have also been tabulated. Clearly, the mass and atomic percentages of desired elements i.e. Gd, Fe, Cr and O in the as-prepared nanoparticles match quite well with the theoretical values which further justifies the successful synthesis of GFCO double perovskite. 

\begin{figure*}[t]
\centering
\includegraphics[width= 160 mm,scale=0.8]{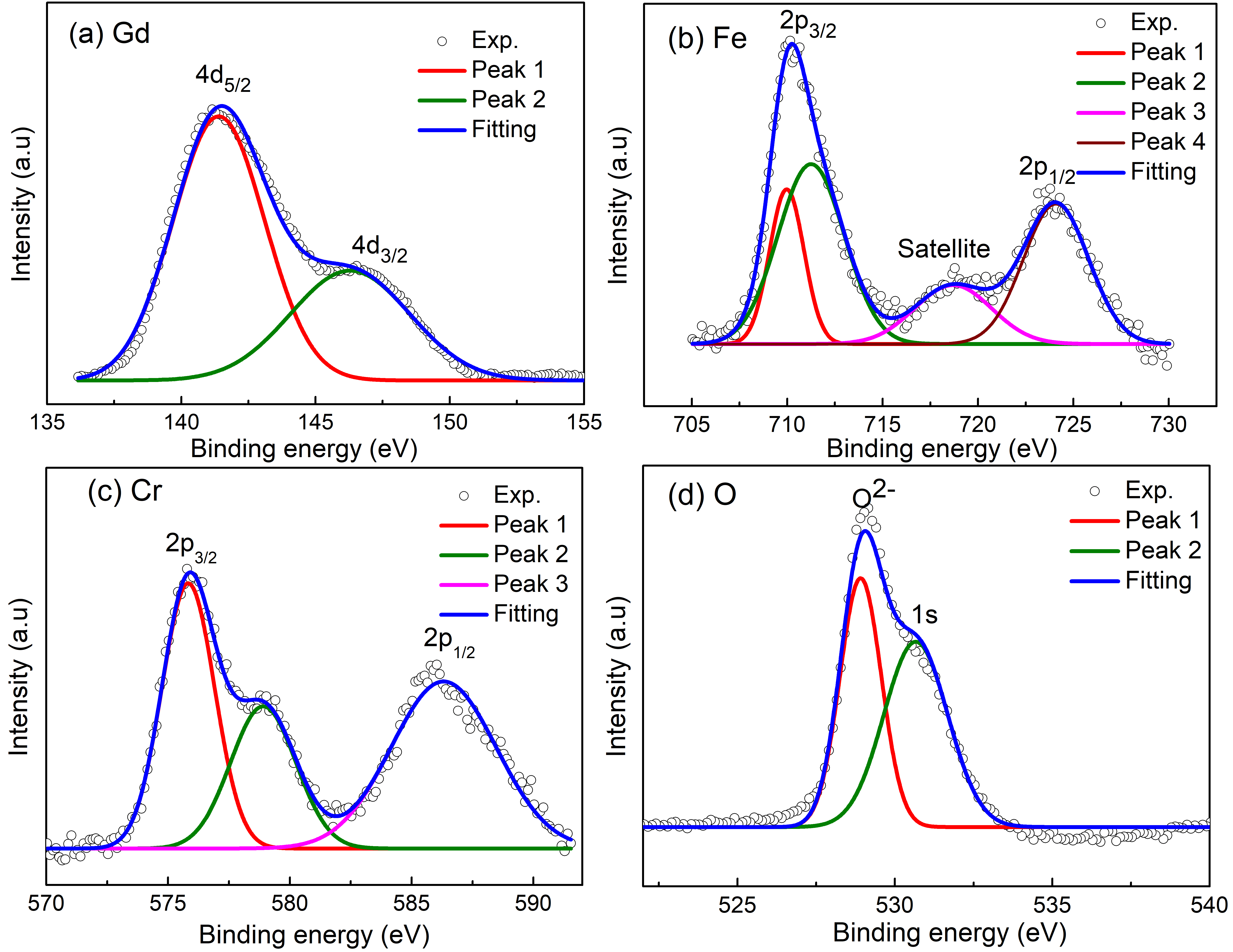}%
\caption{\label{fig:epsart}(a-d) High-resolution XPS core spectra of Gd 4d,
Fe 2p, Cr 2p and O 1s of Gd$_2$FeCrO$_6$ nanoparticles, respectively. The
black circles are indicating the experimental (Exp.) spectra, while the blue solid line is the sum of the fitted peaks. Peak 1, peak 2, peak 3 and peak 4 are sorted by ascending order of binding energy.}
\end{figure*}

Figure 3(a) shows a bright field transmission electron microscopy (TEM) image of GFCO double perovskite. The particle size is in the range of 60-100 nm which is pretty consistent with the values obtained from FESEM imaging. Moreover, we observed clearly that the size of some of the particles are 30-40 nm as shown in the inset of Fig. 3(a). The high resolution TEM (HRTEM) image shown in Fig. 3(b) represents the crystal lattice fringes of nanostructured GFCO with an inter-planar spacing of 0.28 nm corresponding to its (1 2 1) plane.  The inset of Fig. 3(b) provided the selected area electron diffraction pattern of GFCO nanocrystal. 

\subsection{XPS analysis}

The chemical binding energies and oxidation states of different elements of as-synthesized GFCO perovskite were scrutinized using XPS. The high resolution XPS spectra of Gd 4d, Fe 2p, Cr 2p and O 1s core levels for the as-synthesized sample are displayed in Fig. 4(a-d), respectively. The core level XPS spectrum of Gd (Fig.4(a)) can be identified by two characteristic doublet peaks at 4d$_{5/2}$ and 4d$_{3/2}$ corresponding to the binding energy of 141.4 eV and 146.3 eV, respectively \cite{han2017autoclave}. In the XPS spectrum of Fe contents (Fig. 4(b)), the observed two main peaks at 710 eV and 724.08 eV might have arisen due to the spin-orbit splitting of Fe 2p orbital. These two peaks can be assigned to the corresponding states of Fe 2p$_{3/2}$ and Fe 2p$_{1/2}$, respectively and they indicate the Fe$^{3+}$ oxidation state of Fe 2p \cite{das2017pr2fecro6}. Apart from these two peaks, a distinguishable satellite peak can be observed at 718.60 eV. In the case of Fe 2p spectra, we observed two adjacent peaks at 710 eV and 711.20 eV which confirmed the presence of Fe$^{3+}$ and Fe$^{2+}$ oxidation states, respectively \cite{maiti2013large}. The Cr 2p spectrum shown in Fig. 4(c) revealed spin-orbit split 2p$_{3/2}$ and 2p$_{1/2}$ lines which appeared at 575.85 eV and 586.33 eV, respectively. The separation between these two doublets is 10.48 eV which corresponds to the +3 oxidation states of Cr \cite{das2017pr2fecro6}. The initial peak in the binding energy curve of the XPS spectrum of Cr shown in Fig. 4(c) has been deconvoluted into two maxima around 575.85 and 578.92 eV which correspond to the valence states of Cr$^{3+}$ and Cr$^{2+}$, respectively \cite{maiti2013large}. The deconvoluted core-level spectrum (Fig. 4(d)) of oxygen (O$^{2-}$) 1s demonstrated two sets of peak at 528.90 eV due to the singly ionized oxygen atoms and 530.69 eV for the interaction of oxygen atoms within the lattice, respectively.

\subsection{Magnetic characterization}

\begin{figure*}
\centering
\includegraphics[width=160mm]{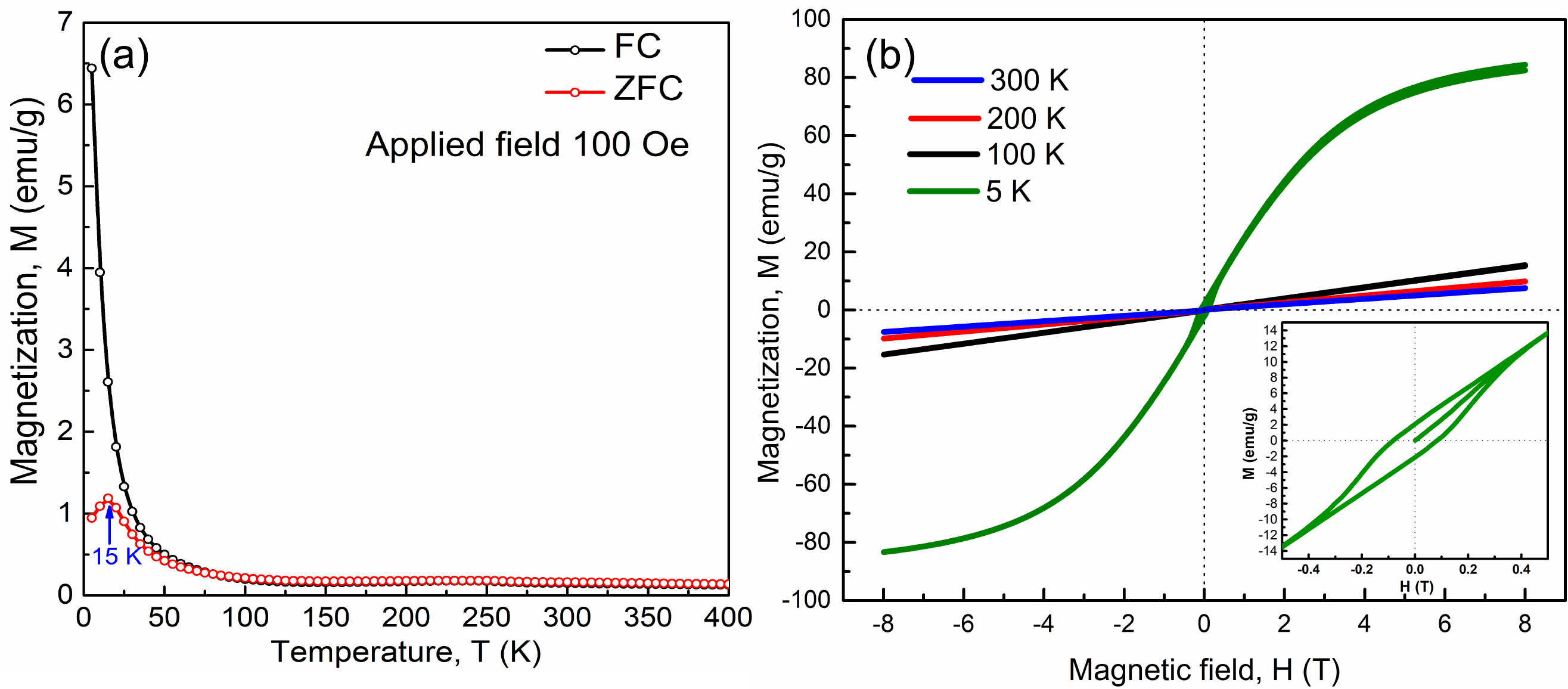}
\caption{\label{fig:epsart} (a) ZFC and FC magnetization vs. temperature measured under 100 Oe applied magnetic field. (b) M-H hysteresis loops recorded at 300 K, 200 K, 100 K and 5 K under an applied magnetic field of up to $\pm$8 T. The inset of (b) demonstrates an enlarged view of the hysteresis loop at 5 K.}
\end{figure*}

Fig. 5(a) displays temperature dependent magnetization (M-T) curves of synthesized GFCO nanoparticles measured via ZFC and FC methods at 5 K to 400 K under 100 Oe external magnetic field \cite{das2020thermal}. As can be observed, the FC magnetization curve continues to increase with decreasing temperature without any sharp transition which indicates the absence of proper long-range order among Fe\textsuperscript{3+} and Cr\textsuperscript{3+} ions as was also observed from XRD analysis. Noticeably, the ZFC and FC curves demonstrate a divergence below 50 K which gradually enlarges at lower temperature below 20 K. Moreover, the ZFC curve exhibits a cusp-like feature at low temperature (T$\sim$15 K) (marked by the blue arrow in Fig. 5(a)) suggesting a N\'eel transition \cite{martin}. Notably, as like as GFCO, a number of analogous double perovskites \cite{martin, cox, primo} have demonstrated N\'eel transition at around 15 K.

Further, field dependent magnetization measurements of synthesized GFCO double perovskite were performed at 300 K, 200 K, 100 K and 5 K under an external magnetic field of up to $\pm$8 T. Fig. 5(b) presents the measured magnetization vs magnetic field (M-H) curves. As can be observed from Fig. 5(b), the M-H curves acquired at 300 K, 200 K and 100 K are linear and not saturated. At these temperatures, no hysteresis loop could be detected indicating the paramagnetic nature of as-prepared GFCO nanoparticles. Inset of Fig. 5(b) provides a magnified view of the magnetic hysteresis loop obtained at 5 K. This hysteresis loop is nearly saturated with a maximum magnetization of $\sim$84 emu/g at an applied magnetic field of 8 T [Fig. 5(b)] which indicates the presence of FM state in GFCO nanoparticles at this lowest temperature. From this hysteresis loop,  the values of remanent magnetization ($M_r$) and coercive field ($H_c$) of GFCO are estimated to be $\sim$3.6 emu/g and $\sim$821 Oe, respectively. Notably, at 5 K this  narrow hysteresis loop was not completely saturated even upon the application of a magnetic field of 8 T (80 kOe) which indicates the presence of AFM states in GFCO nanoparticles along with FM domains  \cite{wu2020structural}. These multiple magnetic domain states is a signature of complex magnetic behavior of GFCO double perovskite at low temperature which can be visualized as follows. From structural analysis, we observed that the Fe\textsuperscript{3+} and
Cr\textsuperscript{3+} ions in GFCO double perovskite were not perfectly ordered at B sites (B and B$^{\prime}$). Moreover, the XPS analysis evinced the presence of mixed valence states of Fe and Cr in GFCO double perovskite. Therefore, in the absence of long-range ordering among Fe\textsuperscript{3+} and Cr\textsuperscript{3+} sites, three next nearest neighbor configurations of B-B$^{\prime}$ ions i.e. Fe\textsuperscript{2+}-Fe\textsuperscript{3+}, Cr\textsuperscript{2+}-Cr\textsuperscript{3+} and Fe\textsuperscript{3+}-Cr\textsuperscript{3+} may exist in GFCO. Among them, the Fe\textsuperscript{2+}-Fe\textsuperscript{3+} and Cr\textsuperscript{2+}-Cr\textsuperscript{3+} pairs will interact via the intervening O\textsuperscript{2-} ions which will give rise to anti-parallel alignment of magnetic spins and eventually generate non-zero spin moment resulting in AFM behavior \cite{das2017pr2fecro6}. Further, the magnetic interaction between Fe\textsuperscript{3+}–Cr\textsuperscript{3+} ions can be explained by the Goodenough-Kanamori rule \cite{kanamori1959superexchange}. According to this rule, the 180${}^\circ$ superexchange interaction between the half-filled Fe\textsuperscript{3+}-d(t$_{2g}^{3}$e$_{g}^{2}$) orbital and empty Cr\textsuperscript{3+}-d(t$_{2g}^{3}$e$_{g}^{0}$) orbital through O\textsuperscript{2-} anion results in FM state. However, the nature of this magnetic interaction transforms into AFM when the d$(e^2_g)$$-$O$-$d$(e^0_g)$ angle gets smaller than 180${}^\circ$ due to lattice distortion \cite{feng2014high}. Especially, for d$^5$$-$d$^3$ systems like GFCO, the superexchange interaction varies from FM to AFM for a range of 125${}^\circ$ to 150${}^\circ$ bond angle \cite{feng2014high}. As demonstrated before, the Fe\textsuperscript{3+}–O\textsuperscript{2-}-Cr\textsuperscript{3+} bond in GFCO double perovskite bends to $\sim$148${}^\circ$ due to octahedral distortion which can also be imputed to the emergence of the competing AFM interaction along with the FM state.

Furthermore, the M–H hysteresis loop at 5 K [inset of Fig. 5(b)] demonstrates an asymmetric shift although nominal along the field axis indicating the presence of intrinsic exchange bias effect in this double perovskite \cite{basith2015tunable}. This is another indication of the concurrence of FM and AFM ordering in as-synthesized GFCO double perovskite \cite{wu2020structural, basith2015tunable}. From the loop asymmetry, we have determined the exchange bias field as $\sim$15 Oe. However, further investigation needs to be performed in ZFC condition to confirm the presence of exchange bias effect in this double perovskite. 

\subsection{Optical characterizations and applications}

\begin{figure}[t]
\centering
\includegraphics[width=70 mm]{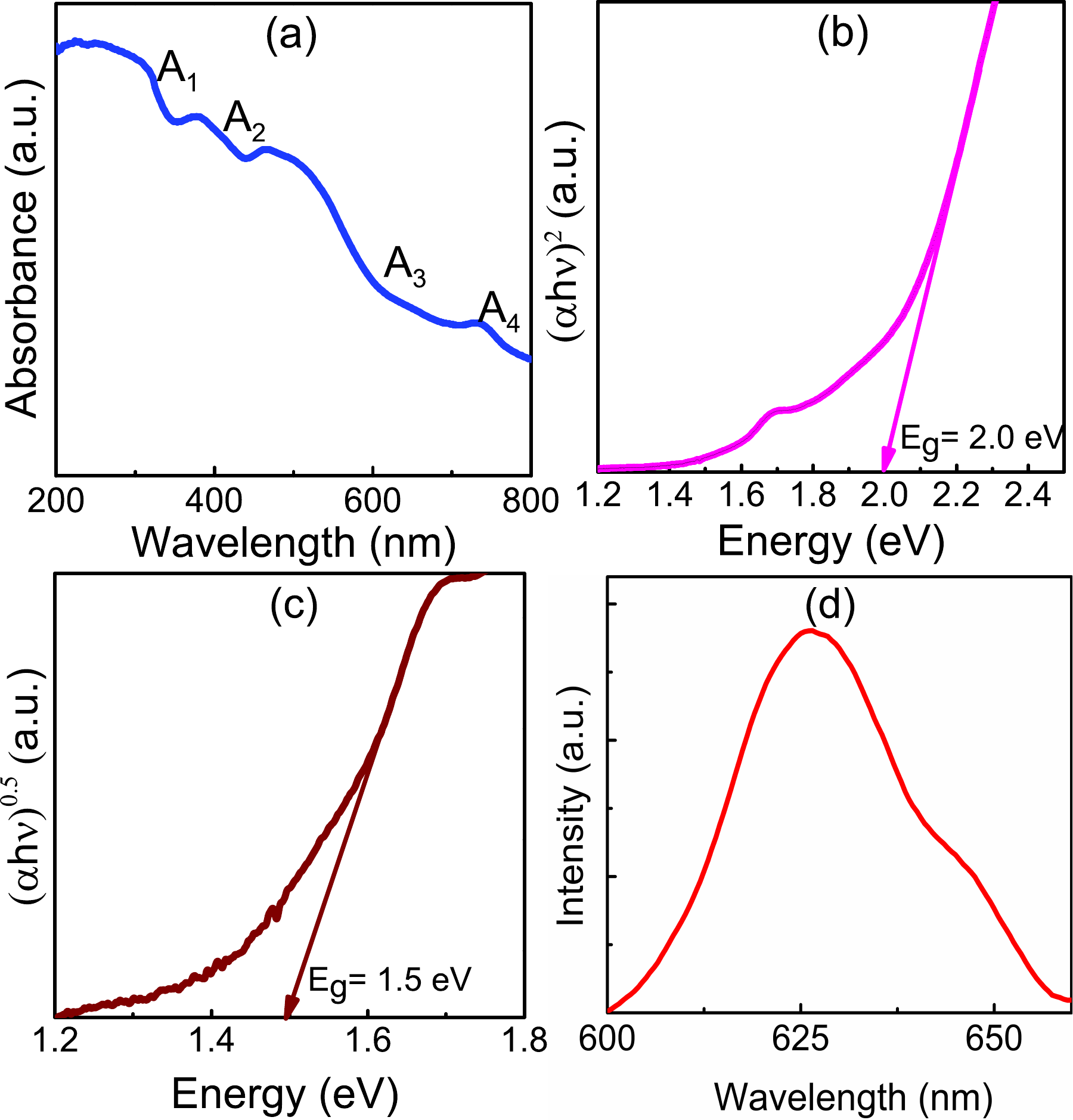}
\caption{\label{fig:epsart} Experimentally obtained (a) optical absorbance spectrum, Tauc plots for direct (b) and (c) indirect optical band gap estimation and (d) steady-state photoluminescence spectrum of Gd$_2$FeCrO$_6$ nanoparticles.}
\end{figure}

The optical characteristics of as-synthesized GFCO nanoparticles were extensively investigated by obtaining their UV-visible absorbance spectrum. As can be seen in Fig. 6(a), the experimentally obtained absorbance spectrum demonstrated four absorption bands (marked as A\textsubscript{1}, A\textsubscript{2}, A\textsubscript{3} and A\textsubscript{4}) both in the UV and visible range indicating the multiband electronic structure of synthesized GFCO nanoparticles \cite{gaikwad2019structural}. The high-energy absorption bands around 333 nm (A\textsubscript{1}) and 412 nm (A\textsubscript{2}) can be attributed to the p-d charge transfer transition [O(2p)$\rightarrow$Fe/Cr(3d)] in Fe/CrO$_6$ octahedral centers of GFCO \cite{arima1993variation}. The absorption band, A\textsubscript{3} around 620 nm is associated with the p-p electronic transitions and the weak absorption band, A\textsubscript{4} around 772 nm might be imputed to the d-d transition in Fe\textsuperscript{3+} \cite{gaikwad2019structural}. \\
Thereafter, we employed the absorbance data to calculate the optical band gap of synthesized GFCO double perovskite. Notably, the optical band gap of GFCO was determined considering both direct and indirect transition using Tauc relation \cite{tauc1966optical} which is given by-


\begin{eqnarray}
\alpha h \nu=A(h\nu-E_g)^n  
\end{eqnarray}

where $\alpha$ is the absorption coefficient, $h\nu$, $A$, $E\textsubscript{g}$ and $n$ signify the energy of photons, proportionality constant, optical band gap and the electronic transition type, respectively. The value of $n$ would be 0.5 for direct transition and 2 for indirect one \cite{nowak2009determination}. Fig. 6(b) and 6(c) demonstrate the generated Tauc plots for estimating the direct and indirect optical band gaps of GFCO nanoparticles, respectively. In these figures, the abscissa intercepts of the tangents to the linear region of the plots provide the values of optical band gaps. As can be observed, for direct and indirect transitions, we have determined the band gap values to be 2.0 and 1.5 eV, respectively.\\
\tab Further, we have obtained the steady-state PL spectrum of as-synthesized GFCO nanoparticles for an excitation wavelength of 230 nm to get insight into their charge-carrier recombination process. As shown in Fig. 6(d), a PL peak was observed at around 626 nm suggesting that upon excitation, the photogenerated electrons and holes recombine radiatively in GFCO double perovskite. Notably, from the position of the PL peak, the band gap value of GFCO is found to be 1.98 eV which closely matches with the direct band gap value as obtained from the Tauc plot (Fig. 6(b)). Moreover, it is worth noting that in a given span of time, radiative recombination is much less likely to occur in indirect band gap materials as compared with the direct ones. This is due to the fact that along with photon, the radiative recombination in indirect band gap materials also requires the involvement of the absorption and emission of phonon. Therefore, it is plausible to infer that our as-synthesized nanostructured GFCO is a direct band-gap double perovskite with a band gap value of $\sim$2.0 eV. Notably, this outcome indicates the semiconducting nature of GFCO and most importantly, reveals its ability to absorb the visible light of the solar spectrum efficiently.\\
\begin{figure}
\centering
\includegraphics[width=80mm]{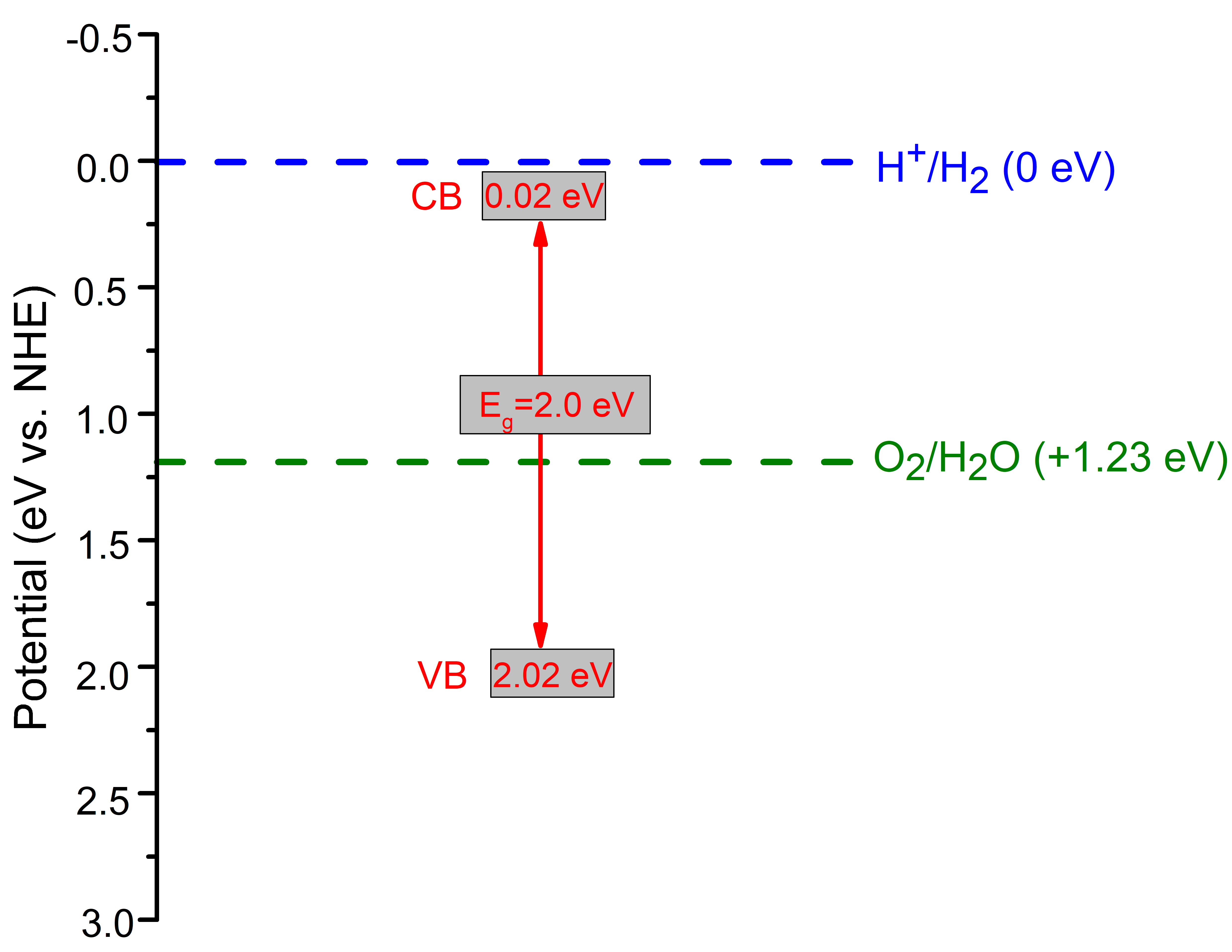}
\caption{\label{fig:epsart} Schematic diagram of the estimated band positions of Gd$_2$FeCrO$_6$ nanoparticles.}
\end{figure}
\tab Moreover, we have determined the band edge positions of as-synthesized GFCO nanoparticles to evaluate their potential for different optical applications. Mulliken electronegativity approach \cite{mulliken1934new,mulliken1935electronic} was adopted to calculate the conduction band minimum (CBM) and valence band maximum (VBM) potentials of GFCO. According to this approach-
\begin{equation}
E_{CBM}=\chi-E_c-\frac{1}{2}E_g
\end{equation}
\begin{equation}
    E_{VBM}=E_{CBM}+E_g
\end{equation}
where $\chi$ is the Mulliken electronegativity of GFCO which is calculated to be 5.52 eV based on the electron affinities and atomic ionization energies. $E\textsubscript{c}$ is the energy of free electrons on the hydrogen scale ($\sim$4.5 eV) and $E\textsubscript{g}$ is the direct optical band gap (2.0 eV) as obtained from the Tauc plot (Fig.  6(b)). Notably, the CBM and VBM potentials (vs. normal hydrogen electrode potential, NHE) of GFCO are calculated to be 0.02 and 2.02 eV, respectively. In order to clearly elucidate the band diagram of GFCO in Fig. 7, we have illustrated the estimated band edge positions of GFCO with respect to the H\textsuperscript{+}/H\textsubscript{2} potential level (0 eV vs NHE, pH 0) and the O\textsubscript{2}/H\textsubscript{2}O potential level (1.23 eV vs NHE, pH 0). Theoretically, if the CBM of a semiconductor is more negative than that of
H\textsuperscript{+}/H\textsubscript{2} potential, photocatalytic H\textsubscript{2} evolution reaction (HER) is likely occur. Similarly, if the VBM is more positive than O\textsubscript{2}/H\textsubscript{2}O potential, O\textsubscript{2} gas will evolve. As shown in Fig. 7, the VBM potential of GFCO is sufficiently positive with respect to the oxidation potential of water which unveils the promising potential of GFCO nanoparticles to be employed in photocatalytic O$_2$ evolution \cite{reunchan2016electronic}. Notably, it is challenging to find metal oxide semiconductors which can enable photocatalytic decomposition of water under visible light illumination. Hence, the favorable surface morphology, optimal direct band gap of 2.0 eV and the band edge positions of GFCO double perovskite make it a promising candidate for visible light responsive photocatalysis, especially for oxygen evolution from water.

\section{Conclusions}

We have synthesized Gd$_2$FeCrO$_6$ (GFCO) double perovskite  via sol-gel citrate method and extensively investigated its structural, magnetic and optical properties. The Rietveld refined powder XRD pattern and electron microscopy imaging demonstrated the successful synthesis of GFCO nanoparticles with $\sim$70 nm average particle size. Thermogravimetric analysis confirmed excellent thermal stability of the synthesized nanoparticles up to a temperature of 1000 $^{\circ}$C. During field cooling, the increase of magnetization with decreasing temperature without any sharp transition indicated the absence of proper long-range order among Fe\textsuperscript{3+} and Cr\textsuperscript{3+} ions as was also suggested by XRD and XPS analysis. As a consequence, at 5 K, we observed the coexistence of FM and AFM states in GFCO double perovskite. This was further evident from the emergence of intrinsic exchange bias effect at 5 K. Both UV-visible and photoluminescence spectroscopic analysis demonstrated  that  band  gap  of  GFCO  perovskite  is within  visible  range  and therefore,  it can  be  concluded  that  this  synthesized  direct  band  gap  semiconductor might have the ability to function as a photocatalyst material. Moreover, the band edge positions of as-synthesized GFCO nanoparticles revealed that the valence band maximum potential is sufficiently positive with respect to the oxidation potential of water which unveils the promising potential of these nanoparticles to be employed in photocatalytic oxygen evolution from water.

\section*{Acknowledgments}
 The financial assistance from the Committee for Advanced Studies and Research (CASR), Bangladesh University of Engineering and Technology (BUET) is acknowledged. The authors also acknowledge Professor Tadahiro Komeda, Institute of Multidisciplinary Research for Advanced Materials (IMRAM), Tohoku University, Japan for XPS measurements.



\begin{thebibliography}{0}%
\makeatletter
\providecommand \@ifxundefined [1]{%
 \@ifx{#1\undefined}
}%
\providecommand \@ifnum [1]{%
 \ifnum #1\expandafter \@firstoftwo
 \else \expandafter \@secondoftwo
 \fi
}%
\providecommand \@ifx [1]{%
 \ifx #1\expandafter \@firstoftwo
 \else \expandafter \@secondoftwo
 \fi
}%
\providecommand \natexlab [1]{#1}%
\providecommand \enquote  [1]{``#1''}%
\providecommand \bibnamefont  [1]{#1}%
\providecommand \bibfnamefont [1]{#1}%
\providecommand \citenamefont [1]{#1}%
\providecommand \href@noop [0]{\@secondoftwo}%
\providecommand \href [0]{\begingroup \@sanitize@url \@href}%
\providecommand \@href[1]{\@@startlink{#1}\@@href}%
\providecommand \@@href[1]{\endgroup#1\@@endlink}%
\providecommand \@sanitize@url [0]{\catcode `\\12\catcode `\$12\catcode
  `\&12\catcode `\#12\catcode `\^12\catcode `\_12\catcode `\%12\relax}%
\providecommand \@@startlink[1]{}%
\providecommand \@@endlink[0]{}%
\providecommand \url  [0]{\begingroup\@sanitize@url \@url }%
\providecommand \@url [1]{\endgroup\@href {#1}{\urlprefix }}%
\providecommand \urlprefix  [0]{URL }%
\providecommand \Eprint [0]{\href }%
\providecommand \doibase [0]{http://dx.doi.org/}%
\providecommand \selectlanguage [0]{\@gobble}%
\providecommand \bibinfo  [0]{\@secondoftwo}%
\providecommand \bibfield  [0]{\@secondoftwo}%
\providecommand \translation [1]{[#1]}%
\providecommand \BibitemOpen [0]{}%
\providecommand \bibitemStop [0]{}%
\providecommand \bibitemNoStop [0]{.\EOS\space}%
\providecommand \EOS [0]{\spacefactor3000\relax}%
\providecommand \BibitemShut  [1]{\csname bibitem#1\endcsname}%
\let\auto@bib@innerbib\@empty
\end{thebibliography}%


\begin{thebibliography}{00}


\bibitem{lv2019copper/cobalt} Y. Lv,  Z. Li,  Y. Yu,  J. Yin,  K. Song,  B. Yang,  L. Yuan,  X. Hu,  Copper/Cobalt-doped LaMnO\textsubscript{3} perovskite oxide as a bifunctional catalyst for rechargeable Li-O\textsubscript{2} batteries, J. Alloys Compd. 801 (2019) 19–26. https://doi.org/10.1016/j.jallcom.2019.06.114. 
\bibitem{cohen1992origin} R. E. Cohen, Origin of ferroelectricity in perovskite oxides, Nature 358 (1992) 136-138. https://doi.org/10.1038/358136a0.
\bibitem{bhalla2000perovskite} A. S. Bhalla, R. Guo, R. Roy, The perovskite structure-a review of its role in ceramic science and technology, Mater. Res. Innov. 4 (2000) 3-26. https://doi.org/10.1007/s100190000062.
\bibitem{basith2017preparation} M. A. Basith, M. A. Islam, B. Ahmmad, M. D. S. Hossain, K. M{\o}lhave, Preparation of high crystalline nanoparticles of rare-earth based complex pervoskites and comparison of their structural and magnetic properties with bulk counterparts, Mater. Res. Express 4 (2017) 075012. https://doi.org/10.1088/2053-1591/aa769e.
\bibitem{basith2014room} M. A. Basith, O. Kurni, M. S. Alam, B. L. Sinha, B. Ahmmad, Room temperature dielectric and magnetic properties of Gd and Ti co-doped BiFeO\textsubscript{3} ceramics, J. Appl. Phys. 115 (2014) 024102. https://doi.org/10.1063/1.4861151.
\bibitem{vasala2015a2b} S. Vasala, M. Karppinen, A\textsubscript{2}B$^\prime$ B$^{\prime\prime}$O\textsubscript{6} perovskites: a review, Prog. Solid. State Ch. 43 (2015) 1-36. https://doi.org/10.1016/j.progsolidstchem.2014.08.001.
\bibitem{gray2010local} B. Gray, H. N. Lee, J. Liu, J. Chakhalian, J. W. Freeland, Local electronic and magnetic studies of an artificial La\textsubscript{2}FeCrO\textsubscript{6} double perovskite, Appl. Phys. Lett.  97 (2010) 013105. https://doi.org/10.1063/1.3455323.
\bibitem{das2017pr2fecro6} N. Das, S. Singh, A. G. Joshi, M. Thirumal, V. R. Reddy, L. C. Gupta, A. K. Ganguli, Pr\textsubscript{2}FeCrO\textsubscript{6}: a type I multiferroic, Inorg. Chem. 56 (2017) 12712-12718. https://doi.org/10.1021/acs.inorgchem.7b01086.
\bibitem{das2008electronic} H. Das, U. V. Waghmare, T. Saha-Dasgupta, D. D. Sarma, Electronic structure, phonons, and dielectric anomaly in
ferromagnetic insulating double pervoskite La\textsubscript{2}NiMnO\textsubscript{6}, Phys. Rev. Lett. 100 (2008) 186402. https://doi.org/10.1103/PhysRevLett.100.186402.
\bibitem{bull2003determination} C. L. Bull, D. Gleeson, K. S. Knight, Determination of B-site ordering and structural transformations in the mixed transition metal perovskites La\textsubscript{2}CoMnO\textsubscript{6} and La\textsubscript{2}NiMnO\textsubscript{6}, J. Phys. Condens. Matter 15 (2003) 4927. https://doi.org/10.1088/0953-8984/15/29/304.
\bibitem{yadav2015magnetic} R. Yadav, S. Elizabeth, Magnetic frustration and dielectric relaxation in insulating Nd\textsubscript{2}NiMnO\textsubscript{6} double perovskites, J. Appl. Phys. 117 (2015) 053902. https://doi.org/10.1063/1.4906989.
\bibitem{costilla2020double} S. U. Costilla-Aguilar, M. J. Escudero, R. F. Cienfuegos-Pelaes, J. A. Aguilar-Mart{\'\i}nez, Double perovskite
La\textsubscript{1.8}Sr\textsubscript{0.2}CoFeO\textsubscript{5+$\delta$} as a cathode material for intermediate temperature solid oxide fuel cells, J. Alloys Compd. 862 (2020) 158025. https://doi.org/10.1016/j.jallcom.2020.158025.
\bibitem{masud2015occurrence} M. G. Masud, K. Dey, A. Ghosh, S. Majumdar, S. Giri, Occurrence of magnetoelectric effect correlated to the Dy order
in Dy\textsubscript{2}NiMnO\textsubscript{6} double perovskite, J. Appl. Phys. 118 (2015) 064104. https://doi.org/10.1063/1.4928467.
\bibitem{du2010magnetic} Y. Du, Z. X. Cheng, S. X. Dou, X. L. Wang, H. Y. Zhao, H. Kimura, Magnetic properties of Bi\textsubscript{2}FeMnO\textsubscript{6}: A multiferroic
material with double-perovskite structure, Appl. Phys. Lett. 97 (2010) 122502. https://doi.org/10.1063/1.3490221.
\bibitem{palakkal2015observation} J. P. Palakkal, P. N. Lekshmi, S. Thomas, K. G. Suresh, M. R. Varma, Observation of high-temperature magnetic transition and existence of ferromagnetic short-range correlations above transition in double perovskite La\textsubscript{2}FeMnO\textsubscript{6}, RSC Adv. 5 (2015) 105531-105536. https://doi.org/10.1039/C5RA24092A.
\bibitem{basith2007temperature} M. A. Basith, S. M. Hoque, M. Shahparan, M. A. Hakim, M. Huq, Temperature features of magnetoresistance of layered
manganite La\textsubscript{2}Sm\textsubscript{0.4}Sr\textsubscript{0.6}Mn\textsubscript{2}O\textsubscript{7}, Physica B Condens. 395 (2007) 126-129. https://doi.org/10.1016/j.physb.2007.03.006.
\bibitem{das2019enhanced} S. Das, I. Sultana, M. D. I. Bhuyan, M. A. Basith, Enhanced magnetic softness of double-layered perovskite manganite La\textsubscript{1.7}Gd\textsubscript{0.3}SrMn\textsubscript{2}O\textsubscript{7}, IEEE Magn. Lett. 10 (2019) 1-4. https://doi.org/10.1109/LMAG.2019.2915620.
\bibitem{kobayashi1998room} K. I. Kobayashi, T. Kimura, H. Sawada, K. Terakura, Y. Tokura, Room-temperature magnetoresistance in an oxide material with an ordered double-perovskite structure, Nature 395 (1998) 677-680. https://doi.org/10.1038/27167.
\bibitem{lv2016structural} M. Lv, Y. Wang, L. Lu, R. Wang, S. Ni, G. Liu, X. Xu, Structural dependence of the photocatalytic properties of double
perovskite compounds A\textsubscript{2}InTaO\textsubscript{6} (A= Sr or Ba) doped with nickel, Phys. Chem. Chem. Phys. 18 (2016) 21491-21499. https://doi.org/10.1039/C6CP03522A.
\bibitem{idris2020novel} A. M. Idris, T. Liu, J. H. Shah, X. Zhang, C. Ma, A. S. Malik, A. Jin, S. Rasheed, Y. Sun, C. Li, et al., A novel double perovskite oxide semiconductor Sr\textsubscript{2}CoWO\textsubscript{6} as bifunctional photocatalyst for photocatalytic oxygen and hydrogen evolution reactions from water under visible light irradiation, Sol. RRL 4 (2020) 1900456.  https://doi.org/10.1002/solr.201900456.
\bibitem{halder2019investigating} S. Halder, M. S. Sheikh, R. Maity, B. Ghosh, T. P. Sinha, Investigating the optical, photosensitivity and photocatalytic properties of double perovskites A\textsubscript{2}LuTaO\textsubscript{6} (A= Ba, Sr): A combined experimental and density functional theory study, Ceram. Int. 45 (2019) 15496-15504. https://doi.org/10.1016/j.ceramint.2019.05.053.
\bibitem{shirazi20202} P. Shirazi, M. Rahbar, M. Behpour, M. Ashraf, La\textsubscript{2}MnTiO\textsubscript{6} double perovskite nanostructures as highly efficient visible light photocatalysts, New J. Chem. 44 (2020) 231-238. https://doi.org/10.1039/C9NJ04932K.
\bibitem{miura2001electronic} K. Miura, K. Terakura, Electronic and magnetic properties of La\textsubscript{2}FeCrO\textsubscript{6}: Superexchange interaction for a d$^5$-d$^3$ system, Phys. Rev. B 63 (2001) 104402. https://doi.org/10.1103/PhysRevB.63.104402.
\bibitem{wu2020structural} H. Wu, Z. Pei, W. Xia, Y. Lu, K. Leng, X. Zhu, Structural, magnetic, dielectric and optical properties of double-perovskite Bi\textsubscript{2}FeCrO\textsubscript{6} ceramics synthesized under high pressure, J. Alloys Compd. 819 (2020) 153007. https://doi.org/10.1016/j.jallcom.2019.153007.
\bibitem{kanamori1959superexchange} J. Kanamori, Superexchange interaction and symmetry properties of electron orbitals, J. Phys. Chem. Solids 10 (1959) 87-98. https://doi.org/10.1016/0022-3697(59)90061-7.
\bibitem{goodenough1955theory} J. B. Goodenough, Theory of the role of covalence in the perovskite-type manganites [La,M(II)]MnO\textsubscript{3}, Phys. Rev. 100 (1955) 564. https://doi.org/10.1103/PhysRev.100.564.
\bibitem{ueda1998ferromagnetism} K. Ueda, H. Tabata, T. Kawai, Ferromagnetism in LaFeO\textsubscript{3}-LaCrO\textsubscript{3} superlattices, Science 280 (1998) 1064-1066. https://doi.org/10.1126/science.280.5366.1064.
\bibitem{weinberg1961electron} I. Weinberg, P. Larssen, Electron paramagnetic resonance and antiferromagnetism in LaCrO\textsubscript{3}, Nature 192 (1961) 445-446. https://doi.org/10.1038/192445a0.
\bibitem{peterlin1986antiferromagnetic} T. Peterlin-Neumaier, E. Steichele, Antiferromagnetic structure of LaFeO\textsubscript{3} from high resolution tof neutron diffraction, J. Magn. Magn. Mater. 59 (1986) 351-356. https://doi.org/10.1016/0304-8853(86)90431-2.
\bibitem{ravi2018multiferroism} S. Ravi, Multiferroism in Pr\textsubscript{2}FeCrO\textsubscript{6} perovskite, J. Rare Earths 36 (2018) 1175-1178. https://doi.org/10.1016/j.jre.2018.03.023.
\bibitem{gaikwad2019structural} V. M. Gaikwad, M. Brahma, R. Borah, S. Ravi, Structural, optical and magnetic properties of Pr\textsubscript{2}FeCrO\textsubscript{6} nanoparticles, J. Solid State Chem. 278 (2019) 120903. https://doi.org/10.1016/j.jssc.2019.120903.
\bibitem{iranmanesh2016sol} M. Iranmanesh, M. Lingg, M. Stir, J. Hulliger, Sol gel and ceramic synthesis of Sr$_2$FeMo$_{1-x}$W$_x$O$_6$ (0$\leq$ x$\leq$1) double perovskites series, RSC adv. 6 (2016) 42069-42075. https://doi.org/10.1039/C6RA03923E.
\bibitem{cernea2013preparation} M. Cernea, F. Vasiliu, C. Plapcianu, C. Bartha, I. Mercioniu, I. Pasuk, R. Lowndes, R. Trusca, G. V. Aldica, L. Pintilie, Preparation by sol-gel and solid state reaction methods and properties investigation of double perovskite Sr\textsubscript{2}FeMoO\textsubscript{6}, J. Eur. Ceram. Soc. 33 (2013) 2483-2490. https://doi.org/10.1016/j.jeurceramsoc.2013.03.026.
\bibitem{bijelic2020rational} J. Bijelic, A. Stankovic, M. Medvidovic-Kosanovic, B. Markovic, P. Cop, Y. Sun, S. Hajra, M. Sahu, J. Vukmirovic, D. Markovic, et al., Rational sol-gel based synthesis design and magnetic, dielectric and optical properties study of nanocrystalline Sr\textsubscript{3}Co\textsubscript{2}WO\textsubscript{9} triple perovskite, J. Phys. Chem. C 124 (2020) 12794-12807. https://doi.org/10.1021/acs.jpcc.0c02973.
\bibitem{hasan2016saturation} M. Hasan, M. A. Basith, M. A. Zubair, M. S. Hossain, R. Mahbub, M. A. Hakim, M. F. Islam, Saturation magnetization and band gap tuning in BiFeO\textsubscript{3} nanoparticles via co-substitution of Gd and Mn,  J. Alloys Compd. 687 (2016) 701-706. https://doi.org/10.1016/j.jallcom.2016.06.171.
\bibitem{chanda2015structural} S. Chanda, S. Saha, A. Dutta, T. P. Sinha, Structural and transport properties of double perovskite Dy\textsubscript{2}NiMnO\textsubscript{6}, Mater. Res. Bull. 62 (2015) 153-160. https://doi.org/10.1016/j.materresbull.2014.11.021.
\bibitem{rodriguez1990fullprof} J. Rodriguez-Carvajal, FULLPROF: a program for rietveld refinement and pattern matching analysis, in: satellite meeting on powder diffraction of the XV congress of the IUCr, 127 (1990) Toulouse, France:[sn].
\bibitem{basith2015tunable} M. A. Basith, F. A. Khan, B. Ahmmad, S. Kubota, F. Hirose, D. T. Ngo, Q. H. Tran, K. M{\o}lhave, Tunable exchange bias effect in magnetic Bi\textsubscript{0.9}Gd\textsubscript{0.1}Fe\textsubscript{0.9}Ti\textsubscript{0.1}O\textsubscript{3} nanoparticles at temperatures up to 250 k, J. Appl. Phys. 118 (2015) 023901. https://doi.org/10.1063/1.4926424.
\bibitem{lufaso2001prediction} M. W. Lufaso, P. M. Woodward, Prediction of the crystal structures of perovskites using the software program SPuDS, Acta Crystallogr. Sect. B: Struct. Sci. 57 (2001) 725-738. https://doi.org/10.1107/S0108768101015282.
\bibitem{anderson1993b} M. T. Anderson, K. B. Greenwood, G. A. Taylor, K. R. Poeppelmeier, B-cation arrangements in double perovskites, Prog. Solid. State Ch. 22 (1993) 197-233. https://doi.org/10.1016/0079-6786(93)90004-B.
\bibitem{li2018new} Z. Li, Y. Cho, X. Li, X. Li, A. Aimi, Y. Inaguma, J. A. Alonso, M. T. Fernandez-Diaz, J. Yan, M. C. Downer, et al., New mechanism for ferroelectricity in the perovskite Ca\textsubscript{2-x}Mn\textsubscript{x}Ti\textsubscript{2}O\textsubscript{6} synthesized by spark plasma sintering, J. Am. Chem. Soc. 140 (2018) 2214-2220. https://doi.org/10.1021/jacs.7b11219.
\bibitem{yamada2018complementary} I. Yamada, A. Takamatsu, H. Ikeno, Complementary evaluation of structure stability of perovskite oxides using bond valence and density-functional-theory calculations, Sci. Technol. Adv. Mater. 19 (2018) 101-107. https://doi.org/10.1080/14686996.2018.1430449.
\bibitem{ravi2017room} S. Ravi, C. Senthilkumar, Room temperature multiferroicity in a new Ba\textsubscript{2}FeMnO\textsubscript{6} double perovskite material.  Ceram. Int. 43 (2017) 14441-14445. https://doi.org/10.1016/j.ceramint.2017.07.217.
\bibitem{byeon2003high} S. H. Byeon, M. W. Lufaso, J. B. Parise, P. M. Woodward, T. Hansen, High-pressure synthesis and characterization of perovskites with simultaneous ordering of both the A-and B-site cations, CaCu\textsubscript{3}Ga\textsubscript{2}M\textsubscript{2}O\textsubscript{12} (M= Sb, Ta), Chem. Mater. 15 (2003) 3798-3804. https://doi.org/10.1021/cm034318c.
\bibitem{mcsucker1999p} L. McSucker, R. Von Dreele, D. Cox, D. Louer, P scardi, Rietveld refinement guidelines, J. Appl. Cryst. 32 (1999) 36. https://doi.org/10.1107/S0021889898009856.
\bibitem{rietveld1969profile} H. M. Rietveld, A profile refinement method for nuclear and magnetic structures, J. Appl. Cryst. 2 (1969) 65-71. https://doi.org/10.1107/S0021889869006558.
\bibitem{shi2011local} C. Shi, Y. Hao, Z. Hu, Local valence and physical properties of double perovskite Nd\textsubscript{2}NiMnO\textsubscript{6}, J. Phys. D: Appl. Phys. 44 (2011) 245405. https://doi.org/10.1088/0022-3727/44/24/245405.
\bibitem{chanda2016magnetic} S. Chanda, S. Saha, A. Dutta, J. Krishna Murthy, A. Venimadhav, S. Shannigrahi, T. P. Sinha, Magnetic ordering and conduction mechanism of different electroactive regions in Lu\textsubscript{2}NiMnO\textsubscript{6}, J. Appl. Phys. 120 (2016) 134102. https://doi.org/10.1063/1.4963824.
\bibitem{momma2011vesta} K. Momma, F. Izumi, Vesta 3 for three-dimensional visualization of crystal, volumetric and morphology data, J. Appl. Cryst. 44 (2011) 1272-1276. https://doi.org/10.1107/S0021889811038970.
\bibitem{nasir2019role} M. Nasir, S. Kumar, N. Patra, D. Bhattacharya, S. N. Jha, D. R. Basaula, S. Bhatt, M. Khan, S. W. Liu, S. Biring, et al., Role of antisite disorder, rare-earth size, and superexchange angle on band gap, Curie temperature, and magnetization of R\textsubscript{2}NiMnO\textsubscript{6} double perovskites, ACS Appl. Electron. Mater. 1 (2019) 141-153. https://doi.org/10.1021/acsaelm.8b00062.
\bibitem{guo2008influence} H. Z. Guo, J. Burgess, E. Ada, S. Street, A. Gupta, M. N. Iliev, A. J. Kellock, C. Magen, M. Varela, S. J. Pennycook, Influence of defects on structural and magnetic properties of multifunctional La\textsubscript{2}NiMnO\textsubscript{6} thin films, Phys. Rev. B 77 (2008) 174423. https://doi.org/10.1103/PhysRevB.77.174423.
\bibitem{iliev2007raman} M. N. Iliev, M. V. Abrashev, A. P. Litvinchuk, V. G. Hadjiev, H. Guo, A. Gupta, Raman spectroscopy of ordered double perovskite La\textsubscript{2}CoMnO\textsubscript{6} thin films, Phys. Rev. B 75 (2007) 104118. https://doi.org/10.1103/PhysRevB.75.104118.
\bibitem{guo2006growth} H. Guo, J. Burgess, S. Street, A. Gupta, T. G. Calvarese, M. A. Subramanian, Growth of epitaxial thin films of the ordered double perovskite La\textsubscript{2}CoMnO\textsubscript{6} on different substrates, Appl. Phys. Lett. 89 (2006) 022509. https://doi.org/10.1063/1.2221894.
\bibitem{panitz2000raman} J. C. Panitz, J. C. Mayor, B. Grob, W. Durisch, A Raman spectroscopic study of rare earth mixed oxides, J. Alloys Compd. 303 (2000) 340-344. https://doi.org/10.1016/S0925-8388(00)00606-X.
\bibitem{obregon2014improved} S. Obreg{\'o}n Alfaro, G. Col{\'o}n Ib{\'a}{\~n}ez, Improved O\textsubscript{2} evolution from a water splitting reaction over Er$^{3+}$ and Y$^{3+}$ co-doped tetragonal BiVO\textsubscript{4}, Catal. Sci. Technol. 4 (2014) 2042-2050. https://doi.org/10.1039/C4CY00050A.
\bibitem{nakamoto2006infrared} K. Nakamoto, Infrared and Raman spectra of inorganic and coordination compounds, Handbook of Vibrational Spectroscopy, John Wiley \& Sons, 2006. https://doi.org/10.1002/0470027320.s4104.
\bibitem{cowin2015conductivity} P. I. Cowin, R. Lan, C. T. Petit, S. Tao, Conductivity and redox stability of double perovskite oxide SrCaFe\textsubscript{1+x}Mo\textsubscript{1-x}O\textsubscript{6-$\delta$} (x= 0.2, 0.4, 0.6), Mater. Chem. Phys. 168 (2015) 50-57. https://doi.org/10.1016/j.matchemphys.2015.10.056.
\bibitem{ding2016redox} H. Ding, Z. Tao, S. Liu, Y. Yang, A redox-stable direct-methane solid oxide fuel cell (SOFC) with Sr\textsubscript{2}FeNb\textsubscript{0.2}Mo\textsubscript{0.8}O\textsubscript{6-$\delta$} double perovskite as anode material, J. Power Sources 327 (2016) 573-579. https://doi.org/10.1016/j.jpowsour.2016.07.101.

\bibitem{han2017autoclave} W. Han, Z. Wang, Q. Li, H. Liu, Q. Fan, Y. Dong, Q. Kuang, Y. Zhao, Autoclave growth, magnetic, and optical properties of GdB\textsubscript{6} nanowires, J. Solid State Chem. 256 (2017) 53-59. https://doi.org/10.1016/j.jssc.2017.08.026.
\bibitem{maiti2013large} R. P. Maiti, S. Dutta, M. K. Mitra, D. Chakravorty, Large magnetodielectric effect in nanocrystalline double perovskite Y\textsubscript{2}FeCrO\textsubscript{6}, J. Phys. D: Appl. Phys. 46 (2013) 415303. https://doi.org/10.1088/0022-3727/46/41/415303.

\bibitem{das2020thermal} S. Das, B. Ahmmad, M. A. Basith, Thermal stability of the crystallographic structure of nanocrystalline Nd\textsubscript{0.7}Sr\textsubscript{0.3}MnO\textsubscript{3} manganite with enhanced magnetic properties, AIP Adv. 10 (2020) 095135. https://doi.org/10.1063/5.0017299.



\bibitem{martin} L. Ortega-San Martin, J. P. Chapman, L. Lezama, J. S{\'a}nchez-Marcos, J. Rodr{\'\i}guez-Fern{\'a}ndez, M.I. Arriortua, T. Rojo, Factors determining the effect of Co (II) in the ordered double perovskite structure: Sr\textsubscript{2}CoTeO\textsubscript{6}, J. Mater. Chem. 15 (2005) 183-193. https://doi.org/10.1039/B413341B.
\bibitem{cox} D. E. Cox, G. Shirane, B. C. Frazer, Neutron‐diffraction study of antiferromagnetic Ba\textsubscript{2}CoWO\textsubscript{6} and Ba\textsubscript{2}NiWO\textsubscript{6}, J. Appl. Phys. 38 (1967) 1459-1460. https://doi.org/10.1063/1.1709666. 
\bibitem{primo} V. Primo-Martın, M. Jansen, Synthesis, structure, and physical properties of cobalt perovskites: Sr\textsubscript{3}CoSb\textsubscript{2}O\textsubscript{9} and Sr\textsubscript{2}CoSbO\textsubscript{6-$\delta$}, J. Solid State Chem. 157 (2001) 76-85. https://doi.org/10.1006/jssc.2000.9041.















\bibitem{feng2014high} H. L. Feng, M. Arai, Y. Matsushita, Y. Tsujimoto, Y. Guo, C. I. Sathish, X. Wang, Y. H. Yuan, M. Tanaka, K. Yamaura, High-temperature ferrimagnetism driven by lattice distortion in double perovskite Ca$_2$FeOsO$_6$, J. Am. Chem. Soc. 136 (2014) 3326-3329. https://doi.org/10.1021/ja411713q.
\bibitem{arima1993variation} T. Arima, Y. Tokura, J. Torrance, Variation of optical gaps in perovskite-type 3d transition-metal oxides, Phys. Rev. B 48 (1993) 17006. https://doi.org/10.1103/PhysRevB.48.17006.
\bibitem{tauc1966optical} J. Tauc, R. Grigorovici, A. Vancu, Optical properties and electronic structure of amorphous germanium, Phys. Status Solidi B 15 (1966) 627-637. https://doi.org/10.1002/pssb.19660150224.
\bibitem{nowak2009determination} M. Nowak, B. Kauch, P. Szperlich, Determination of energy band gap of nanocrystalline SbSI using diffuse reflectance spectroscopy, Rev. Sci. Instrum.  80 (2009) 046107. https://doi.org/10.1063/1.3103603.
\bibitem{mulliken1934new} R. S. Mulliken, A new electroaffinity scale; together with data on valence states and on valence ionization potentials and electron affinities, J. Chem. Phys. 2 (1934) 782-793. https://doi.org/10.1063/1.1749394.
\bibitem{mulliken1935electronic} R. S. Mulliken, Electronic structures of molecules XI. electroaffinity, molecular orbitals and dipole moments, J. Chem. Phys. 3 (1935) 573-585. https://doi.org/10.1063/1.1749731.
\bibitem{reunchan2016electronic} P. Reunchan, A. Boonchun, N. Umezawa, Electronic properties of highly-active Ag$_3$AsO$_4$ photocatalyst and its band gap modulation: an insight from hybrid-density functional calculations,  Phys. Chem. Chem. Phys. 18 (2016) 23407-23411. https://doi.org/10.1039/C6CP03633C.








\end{thebibliography}


\end{document}